\documentclass[useAMS,usenatbib,manuscript]{mn2e}
\usepackage{graphicx}
\usepackage{amsmath}
\usepackage{amssymb}
\usepackage{natbib}

\newcommand{\apj}{\textit{ApJ}}
\newcommand{\aj}{\textit{AJ}}

\newcommand{\apjl}{\textit{ApJ Lett.}}\newcommand{\apjs}{\textit{ApJ Suppl.}}

\newcommand{\mnras}{\textit{MNRAS}}

\newcommand{\AsAs}{A\&A}
\newcommand{\LCDM}{$\Lambda$CDM}
\newcommand{\LWDM}{$\Lambda$WDM}

\newcommand{\kms}{{\,\rm km \ s^{-1}}}
\newcommand{\hMpc}{{\ifmmode{h^{-1}{\rm Mpc}}\else{$h^{-1}$Mpc}\fi}}
\newcommand{\hkpc}{{\ifmmode{h^{-1}{\rm kpc}}\else{$h^{-1}$kpc}\fi}}
\newcommand{\hMsun}{{\ifmmode{h^{-1}{\rm {M_{\odot}}}}\else{$h^{-1}{\rm{M_{\odot}}}$}\fi}}
\newcommand{\Msun}{{\ifmmode{{\rm {M_{\odot}}}}\else{${\rm{M_{\odot}}}$}\fi}}

\title[]{The sizes of mini-voids in the local universe: an argument in favor of a warm dark matter model?}
\author[A. Tikhonov, S. Gottl\"ober, G. Yepes and Y. Hoffman]
     {A. V. Tikhonov$^{1}$\thanks{E-mail: avtikh@gmail.com},
     S. Gottl\"ober$^{2}$\thanks{E-mail:
    sgottloeber@aip.de}, G. Yepes$^{3}$\thanks{E-mail:
    gustavo.yepes@uam.es} and  Y. Hoffman$^{4}$\thanks{E-mail:yhoffman@huji.ac.il}\\
  $^{1}$Saint Petersburg State University, Russian Federation\\
  $^{2}$Astrophysical Institute Potsdam, An der Sternwarte 16, 
       14482 Potsdam, Germany\\
  $^{3}$Universidad Aut\'onoma de Madrid, Grupo de Astrofisica,
       28049 Madrid, Spain\\
  $^{4}$Racah Institute of Physics, Hebrew University, Jerusalem 91904, Israel}

\begin{document}
\bibliographystyle{mn2e}

\date{}

\pagerange{\pageref{firstpage}--\pageref{lastpage}} \pubyear{2002}

\maketitle

\label{firstpage}

\begin{abstract}

Using high-resolution simulations within the Cold and Warm Dark Matter
models we study the evolution of small scale structure in the Local
Volume, a sphere of 8 Mpc radius around the Local Group. We compare
the observed spectrum of mini-voids in the Local Volume with the
spectrum of mini-voids determined from the simulations.  We show that
the \LWDM\ model can easily explain both the observed spectrum of
mini-voids and the presence of low-mass galaxies observed in the Local
Volume, provided that all haloes with circular velocities greater than
20 km/s host galaxies.  On the contrary within the \LCDM\ model the
distribution of the simulated mini-voids reflects the observed one if
haloes with maximal circular velocities larger than $35 \kms$ host
galaxies. This assumption is in contradiction with observations of
galaxies with circular velocities as low as 20 km/s in our Local
Universe.  A potential problem of the \LWDM\ model could be the late
formation of the haloes in which the gas can be efficiently
photo-evaporated. Thus star formation is suppressed and low-mass
haloes might not host any galaxy at all.

\end{abstract}


\section{Introduction}\label{sec:intro}

At present the standard model of cosmology is a flat Friedmann
universe whose mass-energy content is dominated by a cosmological
constant (the $\Lambda$ term), a Cold Dark Matter (CDM) component and
baryons. This \LCDM\ model is characterized by only a few parameters
which are determined already with high precision. The \LCDM\ model
describes structure formation at large scales very well, however it
fails on small scales: the standard model predicts much more small
scale structure than observed. This concerns the number of dark matter
sub-haloes inside galactic-sized host haloes
\citep{Klypinetal1999,Mooreetal1999} as well as the number of dwarfs
in low density regions of the universe \citep{Peebles2001}.  A better
understanding of the physics of structure formation on small scales,
in particular of the correct modeling of baryonic physics, could solve
these problems \citep{Hoeftetal06,crain07}. However, any non-baryonic
physics that reduces power on small scales compared to the standard
model will also improve the situation. It is well known that warm dark
matter acts in this direction by erasing power at short scales due to
free streaming.
  
Recently \cite{TK08} studied the spectrum of mini-voids in the
distribution of galaxies down to $M_B = -12$ in a sphere of radius
8\,Mpc around the Milky Way (the Local Volume -- LV). They compared
the spectrum of mini-voids obtained in LV-candidates selected in
numerical simulations based on WMAP1 \citep{wmap1} and WMAP3
\citep{wmap3} cosmologies and concluded that it matches the observed
one only if haloes with circular velocities $V_c > 35-40$\,km/s define
the mini-voids. Thus \cite{TK08} conclude that \LCDM\ haloes with $V_c
< 35-40$\,km/s should not host galaxies brighter than $M_B = -12$, in
fairly agreement with theoretical expectations \citep{Hoeftetal06,
Loeb08}. In the LV, however, a substantial number of quite isolated
galaxies has been observed with a total of regular (circular) and
chaotic motions well below $35\kms$ \citep{Begum2008}.  These observed
galaxies point towards the same overabundance problem that the \LCDM\
model has in case of Milky Way satellites.  If haloes with circular
velocities $V_c < 35 $\,km/s would host galaxies their total number is
predicted to be too large compared to data.

In Warm Dark Matter (WDM) cosmologies the number of low-mass haloes
drops down when their mass is below a critical damping scale which
depends on the mass of the WDM particles ( see e.g. \cite{avila},
\cite{Bode00}, \cite{knebe}.  In this work, we will use
high-resolution N-body simulations of different dark matter
candidates, namely CDM and WDM, in order to study the properties of
LVs in these models and compared with the real LV observations.  In
particular, we will compare the void functions in simulated
"LV-candidates" with observations in order to derive possible
conclusions about the nature of dark matter particles.

Unfortunately, little is known about the nature of the dark
matter. One of the promising dark matter candidates is the gravitino
for which three possible scenarios are imaginable: Warm (superWIMP) or
mixed cold/warm or cold gravitino dark matter
\citep{Steffen2006}. Another possible candidate are sterile neutrinos
which exhibit a significant primordial velocity distribution and thus
damp primordial inhomogeneities on small scales
\citep{Abazajian2006b}. Experiments are far away from the scales which
could probe directly the properties of the particles of which the dark
matter is made of,  so that the free streaming length cannot be
predicted. However, the free streaming length can be associated with the
particles mass for which astrophysical constraints can be determined.

There is a number of observational constraints on the mass of the WDM
particles. Using HIRES data, a lower limit of $m_{WDM} = 1.2$keV has
been reported by \cite{viel2008} which increases to 4.0 keV when a
combination of SDSS and HIRES data was used.  From the power spectrum
of SDSS galaxies, a lower limit of 0.11 keV has been reported by
\cite{abazajian} which also increases to 1.7 keV if the SDSS
Ly$\alpha$ forest is considered and further increases to 3.0 keV if
high-resolution Ly$\alpha$ forest data are taken into account. An
independent estimation of lower limits for the mass of the WDM
particle based on QSO lensing observations are in agreement with the
former results \citep{miranda}. Recently, \cite{Viel2008b} have shown
that the properties of the void population seen in the Lyman$\alpha$
forest spectra depend on the underlying power spectrum and, therefore,
these voids may also constrain the properties of the dark matter.  We
have run a suite of WDM simulations with particle masses ranging from
3 to 1 KeV. However, in this paper we will consider only the results
from simulations done with the lighter WDM particles (1 KeV).  Results
from the other simulations will be reported elsewhere.

In this paper we will focus on the comparison between the observed
spectrum of mini-voids in the Local Universe with the simulated ones
in \LCDM\ and \LWDM\ models. The structure of this paper is as
follows, in Section 2 we describe briefly our numerical simulations,
the halo mass functions and discuss the reliability of the subsequent
analysis. In Section 3 we describe the observational data sample used
here. In Section 4 we discuss our void finding algorithm. In Section 5
we present results from \LWDM\ and \LCDM\ simulations in terms of
spatial distribution and dynamics.  We conclude this paper with a
discussion of the results found.

\section{Simulations}
\label{sec:simu}

For our numerical experiments, we have assumed a spatially flat
cosmological model which is compatible with the 3rd year WMAP data
\citep{wmap3} with the parameters $h=0.73$, $\Omega_m = 0.24$,
$\Omega_{bar} = 0.042$, $\Omega_{\Lambda} = 0.76$, the normalization
$\sigma_8 = 0.75$ and the slope $n=0.95$ of the power spectrum.  A
computational box of $64 \hMpc$ on a size was used for all
simulations. For \LCDM\ a random realization of the initial power
spectrum computed from a Boltzmann code by W. Hu, was done using
$4096^3$ N-body particles.  After applying the Zeldovich approximation
to this set of particles, we reduced the particles to a total number
of $1024^3$, which corresponds to a mass per particle of $1.6\times
10^7 \hMsun$. We used the original displacement field on the $4096^3$
mesh to derive the initial conditions (i.e positions and velocities)
of the final $1024^3$ particles. In this way, we can increase
the mass resolution of all, or a particular area of the simulation up
to 64 times more using zooming techniques. This is specially useful to
study the problem of {\em fake} haloes in WDM (see below). The same random
phases have been  used in both simulation, so we ensure that   
the same structures will be formed in both models. The initial redshift of the
simulations was set to 50 in all cases.

\begin{figure}
 \begin{center}
  \includegraphics[width=0.5\textwidth]{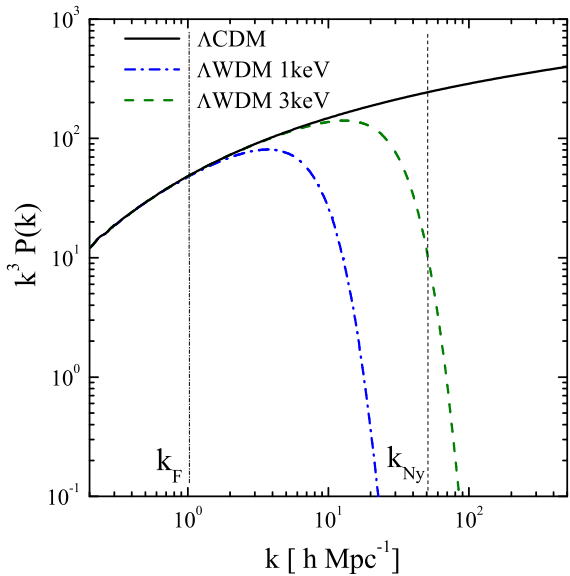}
   \end{center}
   \caption{
     The dimensionless theoretical power spectra for the models
     considered here: CDM (black), and WDM with particle masses: 3 keV
     (blue), and 1 keV (red). The dotted lines correspond to the
    fundamental mode (box size  $64\hMpc$) and the Nyquist frequencies.}
\label{fig:pk} \end{figure}

To generate initial conditions for WDM, we  used  a rescaled version of
the CDM power spectrum  using a fitting function that approximates
the transfer function associated to the free streaming effect
of  WDM particles with $m_ {WDM} = 1-3$ keV \citep{viel05}.  
The power spectra of the three models are shown in Fig.~\ref{fig:pk}.

The effects of the velocity dispersion of the WDM particles have been
neglected in these simulations. They would introduce a certain level
of white noise in the initial conditions that would contribute to the
growth of the small scale power \citep{colin08} that lead to the
formation of spurious halos. At the same time, the rms value of the
random velocity component for 1 kev particles at the starting redshift
of the simulations (z=50) is of the order of 2 km/s which is much
smaller than the typical velocities induced by the gravitational
collapse of the minimum resolved structures in our simulation (see
\citet{zavala} for more details).

These simulations were run with the purpose in mind of studying the
gross structures of the nearby Universe, such as the Local
Supercluster (LSC) and the Virgo cluster, i.e, to use them to do Near
Field Cosmology.  This could be achieved by imposing linear constrains
coming from observations to the otherwise Gaussian random field of
density fluctuations. We used the \cite{HoffmanRibak} algorithm for
imposing the constrains on large scales. A more detailed description
of the simulations and their use for doing Near Field Cosmology will
be reported in a forthcoming paper \citep{yepes09}. However, the
constrained nature of the simulations is ignored here and they are
treated as standard random simulations.

The simulations have been performed using the TreePM parallel N-body
code GADGET2 \citep{Gadget2}. The spatial resolution (Plummer
softening length) was set to the maximum between $1.6 \hkpc$ comoving
and $0.8 \hkpc$ physical at all redshifts.

In an N-body simulation, the computational box size ($k_F$) and the
number of particles ($k_{Ny}$) determine the filtering scales of the
theoretical power spectrum. In Fig.~\ref{fig:pk} we represent these
two scales for our simulations. As can be seen, the Nyquist frequency
is well below the exponential cut-off in the power spectrum for WDM
with $m_{WDM}=3$ keV.  Therefore, we cannot expect too much difference
between the halo mass function of \LCDM\ and this particular \LWDM\
model, since both are dominated mainly by the lack of mass resolution.
Therefore, we decided not to run the 3 keV WDM simulation at this
resolution. Rather, we used it to study the excess of satellites in
Milky Way type haloes, in which re-simulations of haloes with full
resolution (i.e. $4096^3$ particles) have been used for the 3 models
\citep{cairo}. Here we concentrate on the simulation with $1024^3$
WDM-particles of 1 keV mass.

In order to identify haloes in the simulation we have used the AMIGA
Halo Finder (AHF) \citep{ahf}.  The halo positions are first
identified from the local maxima of the density field that is computed
from the particle distribution by Adaptive Mesh Refinement techniques.
The AHF identifies haloes as well as sub-haloes. It provides the virial
masses, virial radii and circular velocities of isolated haloes and
some related quantities for sub-haloes.  We use the maximum circular
velocity $V_c$ to characterize the haloes because the circular
velocity can be easier related to observations than the virial mass.
For reference, haloes with $V_c =50 \kms$ have a virial mass of about
$10^{10}M_{\odot}$ and haloes with $V_c =20 \kms$ have a virial mass
about $10^{9}M_{\odot}$.  We re-scaled all data (coordinates and
masses of haloes) to ``real'' units assuming $H_0 = 72$ km/s/Mpc,
which is the value assumed by \cite{Karachentsev2004} in the observed
LV and which is almost identical with value assumed in the simulations
($H_0=73$ km/s/Mpc)

\begin{figure}
 \begin{center}
  \includegraphics[width=0.5 \textwidth]{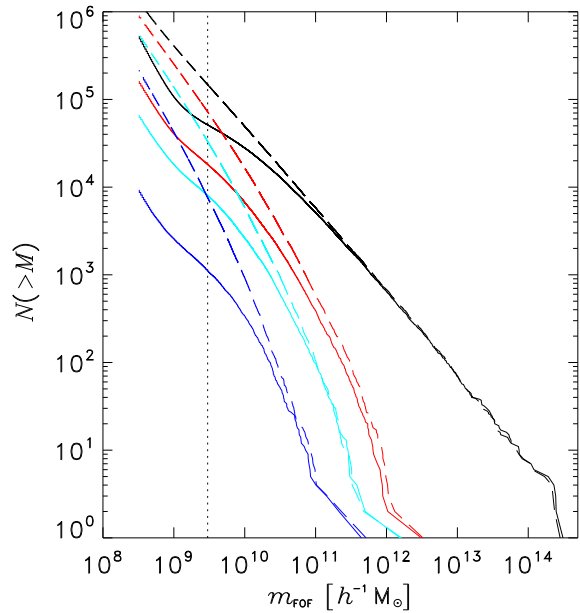}
   \end{center}
    \caption{Cumulative FOF halo mass functions with more than 20
    particles for the \LCDM\ (dashed line) and \LWDM\ (solid line)
    models.  From left to right the  mass functions are shown 
    at redshifts $z = 8$, 6, 5, and 0. The dotted line marks $M_{\rm lim} = 
    3\times10^9 \hMsun$.}  
\label{fig:massfunc} \end{figure}

It is known that in WDM simulations small mass haloes can arise from
the artificial gravitational growth of Poissonian fluctuations that
lead to an unphysical fragmentations of filaments \citep{WW}. These
``fake'' haloes show up typically as tiny haloes regularly aligned
along filaments.  In Figure \ref{fig:pk} one can clearly see the
exponential cut-off of power in the \LWDM\ models. For the 1keV WDM
model the maximum power is reached at $k_{\rm peak} = 3.7 h {\rm
Mpc}^{-1}$. Above this wave number the linear power spectrum contains
very little power so that discreteness noise influences the growth of
small-scale non-linear structures. Below a characteristic mass $M_{\rm
lim}$, which depends on the mass resolution in the simulation, a
spurious fragmentation of filaments occurs which leads to an upturn of
the halo mass function at small scales (Fig.~\ref{fig:massfunc}).
Following \cite{WW} we have estimated for our simulations this
limiting mass as $M_{\rm lim} \approx 3 \times 10^9 \hMsun$.

Figure~\ref{fig:massfunc} shows the evolution of the cumulative mass
functions of halos found by the friend-of-friends algorithm in the
\LCDM\ and \LWDM\ simulations.  The upturn of the \LWDM\ mass function
can be seen at any redshift between $z=8$ and $z=0$ and it is always
at the same position slightly left of the limiting mass. The dotted
line marks the limiting mass $3 \times10^9 \hMsun$ for $z=0$. This
points to a complex process of the formation of spurious
fake-haloes. It seems that they do not form only at at certain epoch
but instead appear all the time after the formation of the first
haloes.  A more detailed numerical study of this problem is
definitively needed, but it is outside of the scope of the present
paper.

Since we see an upturn of the mass function only at $ \approx 1 \times
10^9 \hMsun$ we consider $M_{\rm lim}$ as a rather conservative value.
In terms of circular velocity $M_{upturn}$ corresponds to haloes with
$V_{\rm max} \lesssim 20 {\rm km/s}$ which are candidates to be
spurious. The prime objective of the paper is the study of the
statistical properties of voids, and it is therefore important to
establish that these are not affected by the spurious halos.
Figure~\ref{fig:distr} below proves that this is indeed the case.

\begin{figure*}
 \begin{center}
  \includegraphics[width=0.5 \textwidth,clip]{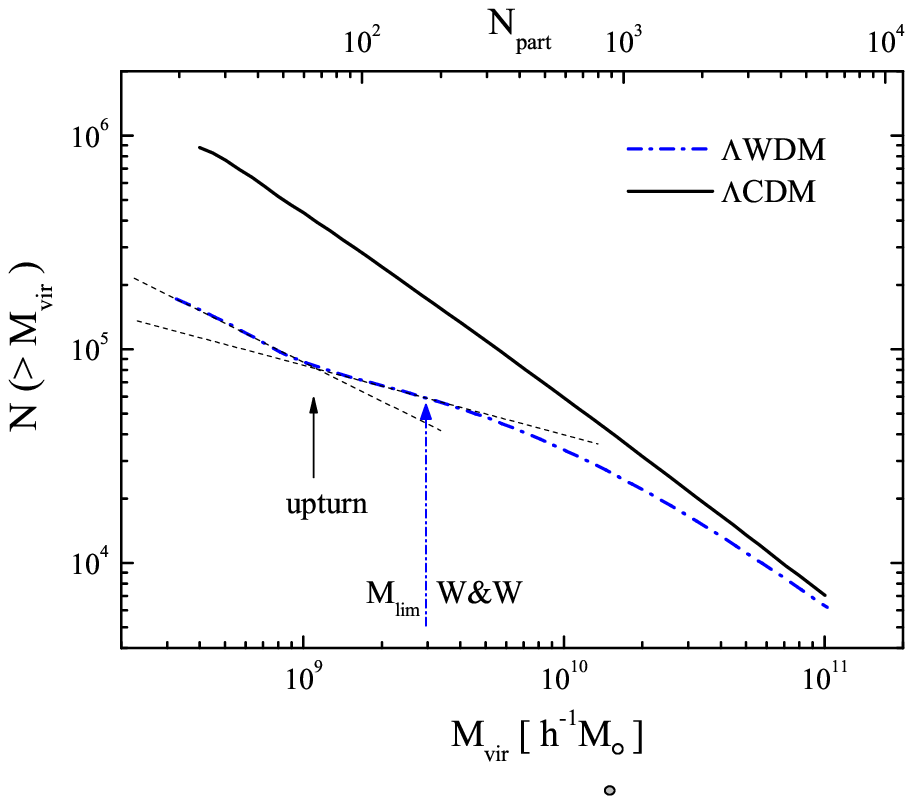}\includegraphics[width=0.5 \textwidth]{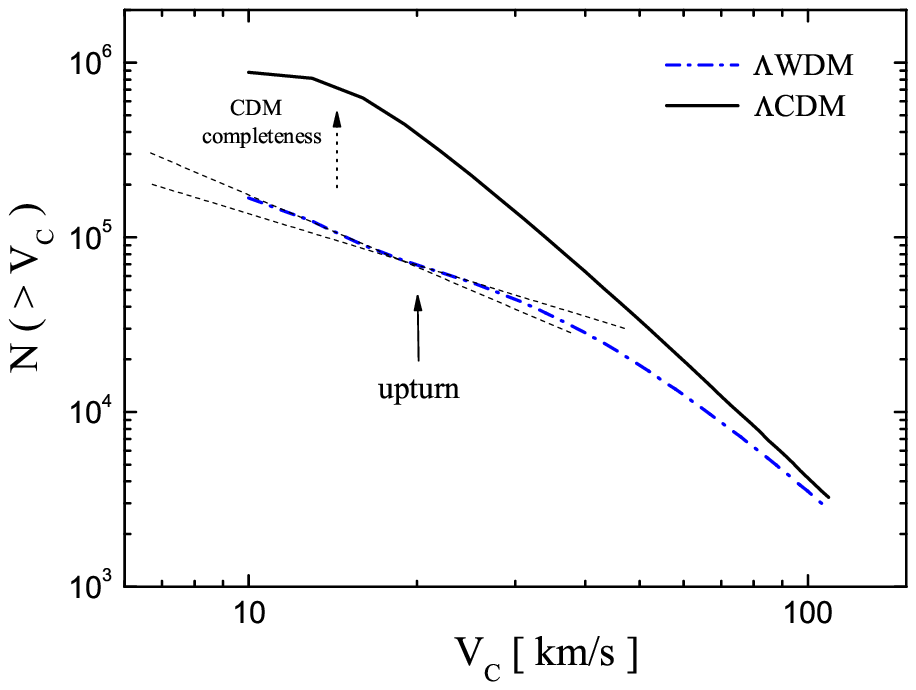}
   \end{center}
    \caption{ 
      The AHF cumulative halo mass functions at redshift $z=0$ (left)
      and the corresponding velocity function (right) for the \LCDM\
      (solid line) and \LWDM\ (dashed-dotted line ) simulations. The
      two dotted lines indicate the
      behavior of the mass function (velocity function) for haloes
      which are very likely fakes and those which should be real ones.
      We mark the intersection point as the mass limit below which
      fake and real haloes are mixed in the WDM simulation. For
      comparison, we also indicate the value for this mass limit
      calculated according to \citet{WW}.  }
\label{fig:mvfahf} \end{figure*}

\section{Observational Data: Local Volume}\label{sec:data}

Over the last decade Hubble Space Telescope observations have provided
distances to many nearby galaxies which where measured with the tip of
the Red Giant Branch (TRGB) stars.  Special searches for new nearby
dwarf galaxies have been undertaken \citet{Karachentsev2004}. Also
special cross-check has been done with IR and blind HI surveys
\citep{Kar07ap}.  The distances were measured from TRGB stars,
Cepheids, the Tully-Fisher relation, and some other secondary distance
indicators. Since the distances of galaxies in the Local Volume are
measured independently of redshifts, we know their true 3D spatial
distribution and their radial velocities. At present, the sample
contains about 550 galaxies with distances less than 10\,Mpc.  A
substantial fraction of the distances have been measured with
accuracies as good as 8-10\% \citep{Karachentsev2004}.

Within the Local Volume dwarf systems have been observed down to
extremely low luminosities. This gives us a unique chance to detect
voids, which may be empty of rather faint galaxies.
\citet{Tikhonov2006} and \citet{TK08} analyzed nearby voids. Here we
use the same observational data as in \citet{TK08} - the updated
Catalog of Neighboring Galaxies \citep[][private
communication]{Karachentsev2004}.  We analyze a volume-limited sample
of galaxies with absolute magnitudes $M_B<-12$ within 8\,Mpc
radius. The overall completeness of the sample has been discussed in
\citet{TK08}.

\subsection{Circular vs. rotational velocity}
\label{sec:vmax}

In order to compare results from an observational galaxy sample with
dissipationless N-body simulations one has to assume certain
hypotheses about how galactic properties are related with their host
dark matter halos and, in particular, how the observed rotational
velocity of a galaxy is related to the circular velocity of its dark
matter halo.

A number of different techniques has been developed to associate
galaxies with simulated DM haloes.  The most well known techniques are
the semi-analytical models (see the review of \citet{Baugh2006}). An
alternative approach is based on the conditional luminosity function
\citep{vandenBosch2004}. Based on the merging history of the DM haloes
and a set of physically motivated recipes the semi-analytical models
place galaxies into the DM haloes. A number of free parameters of
these recipes are calibrated by the observation of galaxies. A dark
matter halo may host a luminous galaxy and a number of
satelittes. Although modern high resolution DM simulations resolve
sub-haloes most of the semi-analytical models do not follow haloes
which are accreted by larger ones but rather model the transformation
of these accreted satellites along their orbits.  Recently,
\cite{Maccio2009} (see also \cite{Koposov2009}) claimed that
semi-analytical models can solve the long standing problem of missing
satellites \citep{Klypinetal1999} within the $\Lambda$CDM
scenario. The basic idea behind this solution of the problem is that
the haloes which host a galaxy of a given measured rotational velocity
are more massive than expected by the direct association of rotational
velocity and the haloes maximum circular velocity. Since more massive
haloes are less frequent the problem of missing satellites is solved.
Thus it is solved at the expense of a predicted large population of
dark satellites the observation of which remains another open problem
in cosmology.

\citet{Stoehr02} were one of the first who argued that due to
gravitational tides the total mass associated with dwarf spheroidals
might be much larger than assumed. On the contrary,
\cite{Kazantzidis04} argue that tidal interaction cannot provide the
mechanism to associate the dwarf spheroidals of the Milky Way with the
massive substructures predicted by the $\Lambda$CDM model.  Recently,
\cite{Penarubbia08} analyzed the dynamics of the stellar systems in
local group dwarf spheroidals and came to the conclusion that the
associated dark matter circular velocity is about three times higher
than the measured velocity dispersion of the stars.

All the previous arguments are mainly based on the assumption that the
small galaxies under consideration are satellites of a more massive
galaxy. However, in the analysis we report in this paper we will
consider isolated dwarf galaxies for which these arguments are not
necessarily valid.

In what follows we argue in favor of a direct association of the
observed rotational velocity of a isolated dwarf galaxy to the maximum
circular velocity of a simulated low mass DM halo. Our arguments are
based on the comparison of the beam size of the observations with
theoretical profiles of the DM haloes.

The \cite{Karachentsev2004} galaxy catalog contains for many of its
galaxies the 21\,cm HI line width at the 50\% level from the maximum 
($W_{50}$) as well as the apparent axial ratio ($b/a$). Using these
informations \cite{TK08} estimated the rotational velocity of the
galaxy as $V_{\rm rot} =W_{50}/2\sqrt{1-(b/a)^2}$.  They argue that a
large fraction of the HI line width is likely produced by random
motion of the order of $10 \kms$ but did not subtract those random
motions. Their conclusions were based on a comparison of this
rotational velocity with the circular velocities of the haloes. We
follow their suppositions and associate the rotational velocity
$V_{\rm rot}$ with the peak circular velocity of the dark matter
halo. In order to give additional arguments in support of this
assumption we compare now circular velocity profiles of dark matter
haloes with the HI radius within which the observations are performed.

\begin{figure} 
 \begin{center}
  \includegraphics[width=0.5 \textwidth,clip]{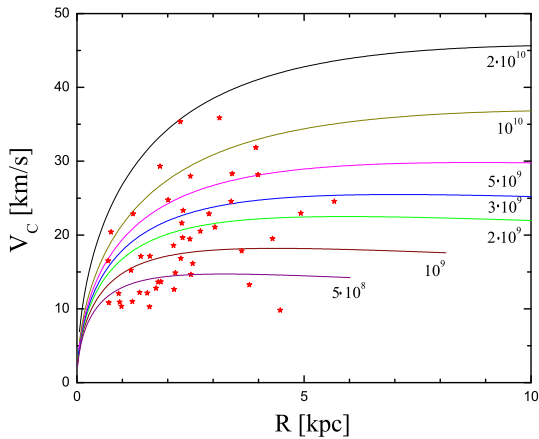}
\end{center} 
   \caption{
     NFW circular velocity profiles for DM haloes with masses between
     $5 \times 10^8 \hMsun$ and $2 \times 10^{10} \hMsun$.  Red stars:
     observed $V_{rot}$ vs HI diameters taken from the FIGGS LV galaxy
     sample.}
\label{fig:Vcmax}  
 
\end{figure}

In Figure~\ref{fig:Vcmax} we plot the NFW velocity profiles up to
$4R_s$ for 7 haloes with masses between $5 \times 10^8 \hMsun$ and $2
\times 10^{10} \hMsun$. Here, we assumed concentrations following the
fitting formula provided by \citet{Neto07} which is in agreement with
\citet{Maccio2007}. Now we want to compare these circular velocities
with the measured rotational velocities of nearby dwarfs. We use the
FIGGS galaxy survey \cite{Begum2008} with $M_B > -14.5$ which is
located inside the Local Volume. In this survey, measurements of
$W_{50}$ together with the HI diameters at a column density of
$\approx 10^{19}$\,atoms cm$^{-2}$ and the inclinations of optical
images ($i_{opt}$) are provided. We estimated the rotational
velocities as $V_{rot} = W_{50}/(2 \times \sin(i_{opt}))$ for galaxies
with $i_{opt} > 40^{\circ}$ and plotted them vs. the HI radii on top
of the NFW profiles in Figure~\ref{fig:Vcmax}.

One can clearly see that most of the observational points are located
on the flattening parts of the NFW velocity profiles with circular
velocities below $30 \kms$.  
Since  dwarf galaxies are expected to be dark
matter dominated,  the association of the observed rotational
velocities at the  HI radius  with the maximum circular velocity of
DM haloes is in our opinion a valid approximation.

\section{Void Finder}\label{sec:voids}

Clusters, filaments and voids constitute the main components of the
large-scale structure of the Universe.  Soon after the discovery of
the first cosmic void in the Bo\"otes constellation \citep{Kirshner81}
it became clear that voids in galaxy distribution are the natural
outcome of gravitational instability \citep{p82,hs82} During the last
two decades the void phenomenon has been discussed extensively both
from a theoretical as well as a observational point of view (see
e.g. \cite{Peebles2001}.  There are many different possibilities to
detect voids. A thorough review and comparison of different algorithms
of void finders has been recently presented in
\cite{2008MNRAS.387..933C}.  The void finders can be classified
according to the method employed. Most are based on the point
distribution of galaxies or DM haloes and some on the smoothed density
or potential fields. Some of the finders are based on spherical
filters while others assume no inherent symmetry an consider voids a
crucial component of the cosmic web (e.g. \cite{Forero2008}).

Our void finder detects empty regions in a point distribution. These
regions are not necessarily spherical.  Our method is not sensitive to
the total number of objects since the objects are mostly located in
high density regions. But it is very sensitive to the appearance of
objects in low density regions. In fact, any object detected in the
middle of a void would divide it into two voids. A detailed
description of the algorithm can be found in \cite{Tikhonov2006}. Our void
finder is somewhat similar to that used by \cite{ElAdPiran97} and
\cite{Gottloeber2003}.  Namely, we place a 3D mesh on the
observational or simulated sample.  On this mesh, the identification
of voids starts with the largest one.  At first, we search for the
largest empty sphere within the considered volume. The voids are
supposed to be completely inside the geometric boundaries of the
sample, therefore, for a given grid point the radius of the largest
empty sphere is determined by the minimum of the distance to the
nearest galaxy (or halo) and the distance to the boundary of the
sample.

The consecutive search for empty seed spheres with decreasing radius,
$R_{\rm seed}$, is combined with their subsequent expansion by means of
addition of spheres whose centers are located inside the given void
and whose radius is $R_{1} > k \cdot R_{\rm seed}$ where we assume
$k=0.9$.  Thus, the spheres added to the void intersect with it by
more than 30\% of the volume. At any search step for an empty seed
sphere, all previously identified voids are taken into account, i.e.
void overlapping is not allowed. The search for voids continues as
long as $R_{\rm seed}$ is larger than a threshold of 0.5 Mpc. This
algorithm is simple, flexible and appropriate for our definition of
voids as regions completely free from galaxies of certain
luminosities.

With our assumption, $k=0.9$, the voids adequately represent the empty
regions between galaxies and at the same time preserve a sufficiently
regular (elliptical) form.  For $k=1$, voids would be strictly
spherical. When $k$ decreases, the voids begin to penetrate into
smaller and smaller holes, and the shapes of voids become more
irregular. Assuming a very small $k$, one single void (the first one)
would fill most part of the sampled volume. The value of the
coefficient $k=0.9$ used in this paper is a compromise, chosen after
empirical detection of voids in distributions of points with different
properties. The voids turn out to be well separated one from another
and they can be approximated by triaxial ellipsoids.

Once we have identified the voids, we calculate the cumulative void
function (VF) of their volumes. We applied also a spherical variant
($k=1$) of our void finder and found that the assumption of sphericity
of voids does not affect our results.

\section{Results}\label{sec:results}

\subsection{Local Volumes in the simulations}

In the following we use the same criteria as \citet{TK08} to select
``LV-candidates'' from our simulations.  These are spheres of radius
8\,Mpc which are characterized by the main features of the observed
Local Volume galaxy sample. In our simulations the LV-candidates must
be centered on a DM-halo with  virial mass in the range
$1\times10^{12}M_{\odot} < m_{vir} < 3\times 10^{12}M_{\odot}$.
This is our definition of a Local Group candidate. We do not require
that this local group consists of two haloes comparable in mass
because the appearance of two haloes instead of one halo of the same
total mass does not change the dynamics of matter outside the Local
group. We also do not require that the Local group should be at the
same distance from a Virgo-type cluster as the observed one.  Instead,
we impose an overdensity constraints on the distribution of mass in
the LV, which we take from the observed LV sample of galaxies.  Below
we summarize the conditions, which were used to find Local volumes in
our simulations:

\begin{enumerate}
\item We put a sphere of radius 8\,Mpc on a Local group candidate.
\item No haloes with $m_{vir}> 2\times 10^{13}M_{\odot}$ are inside
  this sphere.
\item There are no haloes more massive than $5.0\times
  10^{11}M_{\odot}$ in a distance between 1 and 3~Mpc.
\item The centers of the spheres (the Local Group candidates) are
  located at distances larger than 5~Mpc one from the other.
\item The number density of haloes with $V_c>100 \kms$ inside this
  sphere exceeds the mean value in the whole box by a factor in the
  range $1.4-1.8$.
\end{enumerate}

We found 14 LV-candidates in the \LWDM\ simulation which fulfill these
criteria and 9 in the \LCDM\ simulation.  The reason for the
difference is the tight requirement on the number density of massive
halos. This means that in total between 18 and 23 massive halos must
be in this sphere. Thus small differences between the position of a
halo or its circular velocities in the \LCDM\ and \LWDM\ simulations
may already lead to a violation of this criterion.  Since the large
scale structure in both simulations is quite similar we decided to use
in the \LCDM\ simulation the same LV-candidates as in the \LWDM\
simulation. All these LV-candidates fulfill the first four criteria,
7 fulfill also the overdensity criterion, 5 are 10 \% above or below
(what is equivalent to 2 additional or missing halos with $V_c > 100
\kms$) and two miss 5 halos. However, all these five halos are in the
range $90 \kms < V_c < 100 \kms$. The advantage of our choice is that
we can compare the properties of the same regions with the same large
scale environment in both simulations.

\begin{figure*}
 \begin{center}
  \includegraphics[width=\textwidth]{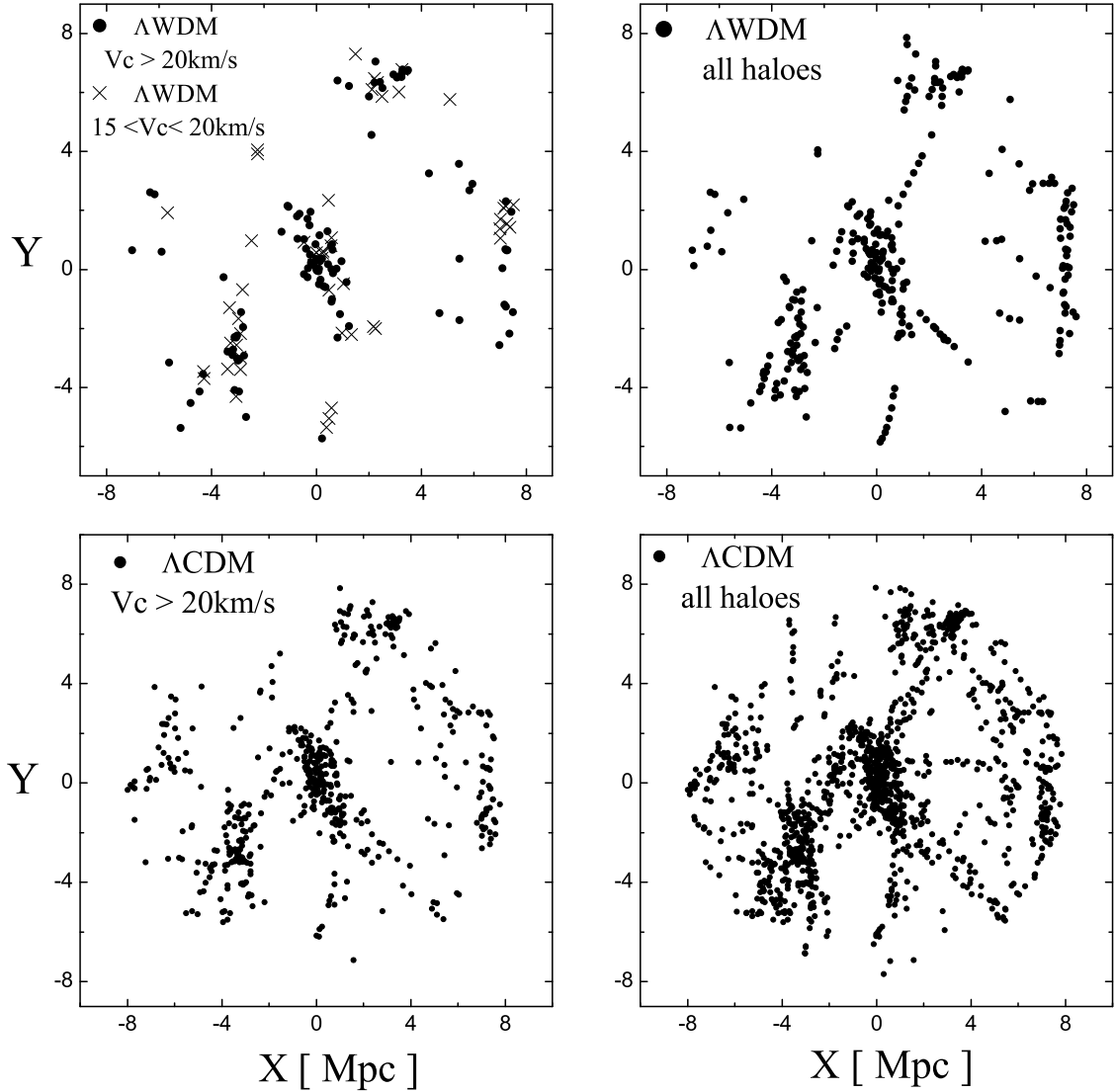}
   \end{center}
   \caption{Distribution of haloes in a slice of 4\,Mpc thickness
     through the center of an LV-candidate.  The upper panel shows
     the \LWDM\ model, on the left plot haloes with $V_c \geq 20\kms$
     are marked by filled circles and haloes with $15 < V_c < 20 \kms$
     by crosses. On the right plot all haloes are shown. The lower
     panel shows the distribution of the corresponding \LCDM\ haloes.}
     \label{fig:distr} 
\end{figure*}

Halo finders like FOF or AHF can find haloes down to 20 particles,
however to determine the internal properties of haloes one needs much
more particles. Moreover, the sample is not necessarily complete down
to 20 particles per halo. There is a large scatter in the $V_c$
vs. $N_{part}$ relation which reduces substantially if one considers
only isolated haloes.  Haloes with $N_{part} \sim 100$ have circular
velocities of $V_c \sim 20$\,km/s. In our simulation this limit
coincides with the upturn of WDM velocity-mass function due to fake
haloes. As an example for the existence of fake haloes in \LWDM\
simulations Figure~\ref{fig:distr} shows the same LV-candidate in both
models: top row for \LWDM\, bottom row for \LCDM\. In the top, right
plot one can clearly see filaments of regularly aligned spurious
fake-haloes which extend across the voids and decrease their sizes.
These are haloes with very low circular velocities. In the top left
plot we have marked haloes with $15 < V_c \leq 20 \kms$ by crosses and
halos with $V_c  >  20 \kms $ by filled circles. One can clearly see
the difference to the right plot: Haloes with $V_c \geq 20 \kms$(filled
circles in the left plot) do not appear at all in artificial
chains. Haloes with $15 < V_c \leq 20 \kms$ only slightly modify the
overall distribution and do not fill voids.  Thus, they do not change
substantially the void statistics.  We did not find any such
artificial regular filament of tiny haloes when we consider only halos
above our limit of reliability $V_c = 20$\,km/s. We conclude that our
analysis is not likely to be affected by spurious haloes as long as we
consider halo samples with circular velocities $V_c > 20 \kms$ and
might be slightly affected for samples with $V_c > 15 \kms$.

\subsection{Properties of voids}\label{sec:cvf}

\begin{figure}
 \begin{center}
  \includegraphics[width=0.5\textwidth]{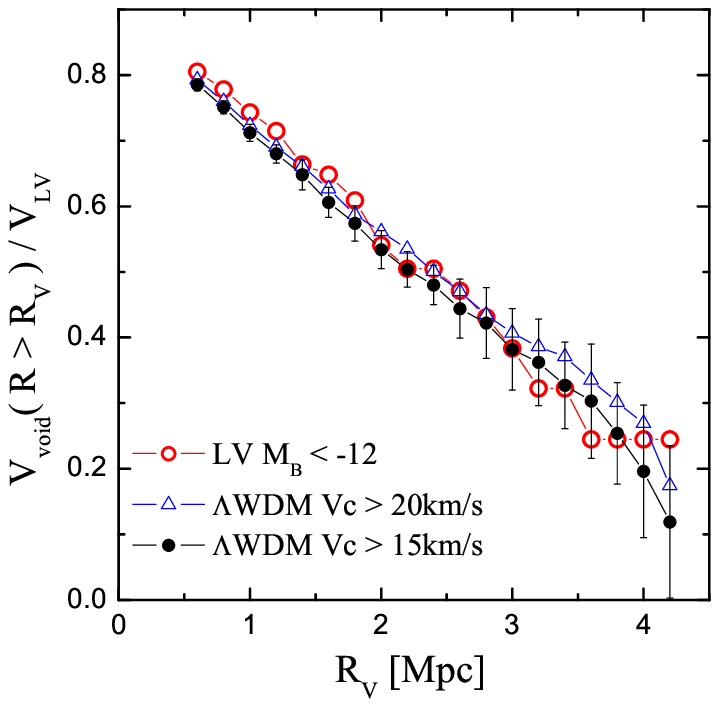}
   \end{center}
   \caption{Volume fraction of the LV occupied my mini-voids
   (VFF). The VFF of the observational sample with $M_B<-12$) (red
   circles) is compared with the mean VFF obtained from the 14 LVs in
   the \LWDM\ simulation with haloes with circular velocity $V_c > 20
   \kms$ (open blue triangles) and $V_c > 15\kms$ (filled black
   circles), for which the $1 \sigma$ scatter is also shown.}
   \label{fig:VFw} 
\end{figure}

\begin{figure}
 \begin{center}
  \includegraphics[width=0.5\textwidth]{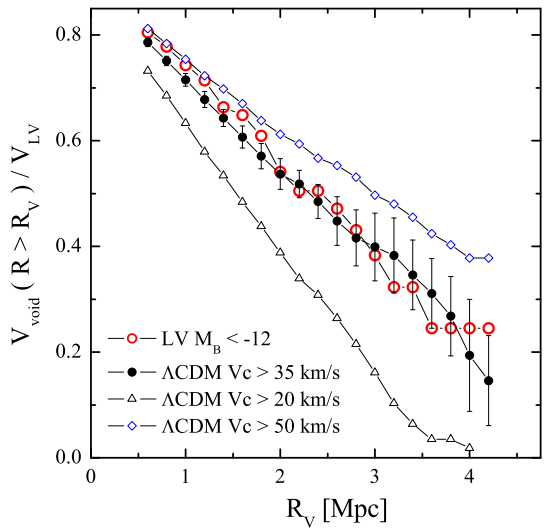}
   \end{center}
   \caption{Volume fraction of the LV occupied my mini-voids
   (VFF). The VFF of the observational sample with $M_B<-12$) (red
   circles) is compared with the mean VFF obtained from the 14 LVs in
   the \LCDM\ simulation with haloes with circular velocity $V_c > 20
   \kms$ (open triangles), $V_c > 50 \kms$ (open diamonds), and $V_c >
   35\kms$ (filled black circles), for which the $1 \sigma$
   scatter is also shown.}  \label{fig:VFc} 
\end{figure}

The Local Volumes contain a large number of mini-voids with radii
$R_{\rm v}$ smaller than about 4 Mpc. We characterize these mini-voids
by the cumulative volume of the mini-voids $V_{\rm void} (R >R_{\rm
V})$ normalized to the Local Volume $V_{\rm {LV}}$. In
Figure~\ref{fig:VFw} we compare the cumulative volume filling fraction
(VFF) of mini-voids ($V_{\rm void} (R >R_{\rm v}) / V_{\rm {LV}}$) of
the 14 LVs found in the \LWDM\ simulation with the observed VFF for
$M_B < -12$ galaxies. The figure indicates that the VFF in the
distribution of haloes with $V_c > 20 \kms$ fits well to the observed
one. In fact, the observed spectrum of local mini-voids is reproduced
within the \LWDM\ model almost over the full range of void volumes.
The size of the large voids reduces slightly if we assume that haloes
with $V_c > 15 \kms $ define the voids. With the present mass
resolution we cannot exclude that this reduction is due to fake
haloes. In the luminosity range $-13.2 < M_B < -11.8$ only three field
galaxies have been observed with $V_{rot} < 15 \kms$ (table~3 of
\citet{TK08}). So far, the existence of such very slowly rotating (low
mass) galaxies is not in contradiction with the assumption of WDM.

The VFF obtained from the \LCDM\ simulation matches observations only
if we assume that objects with $V_{c} > 35 \kms $ define the local
mini-voids (Figure~\ref{fig:VFc}). The \LCDM\ VFF is well above the
data, if we assume $V_{c} > 50 \kms $ for the void definition and it
is far below the observational data if we assume $V_{c} > 20 \kms $,
i.e. in this case large mini-voids are missing in the numerical
realizations of local volumes. Thus, our \LCDM\ simulation is in
agreement with observational data if haloes with $V_{c} < 35 \kms $ do
not host any galaxy. This statement is independent of the volume under
consideration. As a test of reliability we repeated the VFF analysis
assuming a LV of radius 6 Mpc both in observations and in our 14
simulated LVs. We recovered the same circular velocity $V_c \approx 35
\kms $ for haloes hosting galaxies with $M_B<-12$. Thus we are sure
that our original sample is not influenced by voids which intersect
the outer bound of the LV.

However, a significant amount of LV galaxies with magnitudes $M_B \sim
-12$ and rotational velocity as low as $V_{rot} \lesssim 20 \kms$ have
been recently observed. A systematically selected sample of such
extremely faint nearby dwarf irregulars galaxies has been observed
with the Giant Metrewave Radio Telescope \citep{Begum2008}. Dwarf
irregular are typically dark matter dominated. These objects have
surprisingly regular kinematics. One of the faintest ones is
Camelopardalis B with $M_B \sim -10.9$ \citep{Begum2003}.

Within the \LCDM\ model mini-voids with sizes $R_V > 1$ Mpc cover a
volume of about 1600Mpc$^3$ (or about 75 \% of the sample volume, see
Figure~\ref{fig:VFc}). We found a mean of $463 \pm 85$ haloes with $20
< V_c < 35 \kms$ within the mini-voids of a simulated LV where the
scatter is the $1\sigma$ fluctuation between the 14 LVs. This number
is more than an order of magnitude above the observed number of faint
galaxies in the Local Volume. Thus the \LCDM\ model fails to explain
the observed local VFF. The predicted overabundance of low-mass
galaxies in low density regions is similar to another problem well
known in high density regions where the \LCDM\ model predicts an order
of magnitude overabundance of low-mass satellites. Typically the more
massive haloes ($30 < V_c < 35 \kms$) can be found near the borders of
the mini-voids whereas the smallest ones tend to be in the center. A
similar behavior has been found in large voids \citep{Gottloeber2003}.

\subsection{RMS  peculiar velocities}\label{sec:rms}

Next, the Hubble flow around the LG-like objects is studied in an
attempt to confirm that our LV-candidates are indeed the numerical
counterparts of the observed LV. To this end, we calculated the rms
radial velocity deviations from the Hubble flow, $\sigma_H$, both for
the observational data and the objects in the simulated LVs.  As
limits of circular velocities of haloes used in our analysis, we took
the values found from the previous void analysis. We used haloes with
circular velocities $V_c > 20 \kms$ in the \LWDM\ and with $V_c > 35
\kms$ in the \LCDM\ sample. But even if we assume for the \LCDM\
sample the lower value of $V_c > 20 \kms$ the result does not change
substantially in spite of the significant increase in the number of
objects.

\begin{figure}
 \begin{center}
  \includegraphics[width=0.5\textwidth]{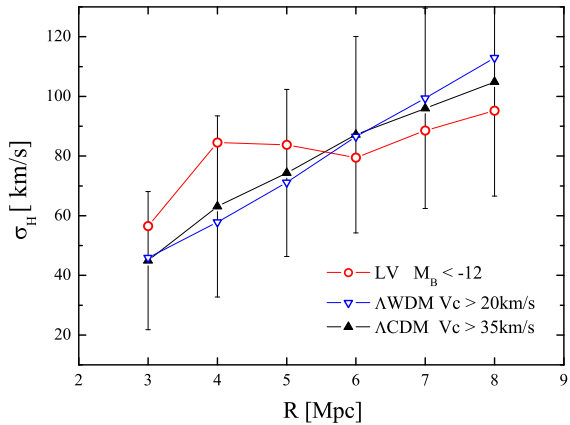}
   \end{center}
   \caption{The rms radial velocity deviations from the Hubble flow
     $\sigma_H$ for galaxies with $M_B<-12$ in the Local Volume with
     distances from 1\,Mpc up to R (red curve with open circles). The
     estimates are corrected for the apex motion and for distance
     errors. Black filled triangles with error bars show the mean
     $\sigma_H$ on corresponding scales for 14 CDM LV-candidates for
     haloes with $V_c > 35\kms$ with 1-$\sigma$ scatter. Blue
     triangles - mean $\sigma_H$ for WDM LV-candidates for haloes with
     $V_c > 20 \kms$. }
   \label{fig:sH} 
\end{figure}

Following \cite{TK08} we calculate $\sigma_H$ in the rest frame of the
considered galaxy sample, i.e. we correct for the apex motion.  We
correct the rms velocities of the galaxies also for distance
errors. Depending on radius these corrections reduce the value of
$\sigma_H$ by about 5\% to 15\% (cf. Table 3 of \cite{TK08}).
Figure~\ref{fig:sH} shows the observed $\sigma_H$ in the range from
1~Mpc to $R$ and the mean value obtained around the Local Group in the
center of each of the 14 LVs in the \LCDM\ and \LWDM\ simulations.  In
Fig. \ref{fig:sH} we compare the $\sigma_H$ of our numerical LVs with
the observed one and found a reasonable agreement for both models. We
error bars show the scatter among the 14 LVs of the \LCDM\ model. In
\LWDM\ we found about the same scatter.  The excess seen in the
observed $\sigma_H$ at $R=4$ and 5 Mpc probably reflects the fact that
most of the galaxy groups of the LV are located within a distance of 3
to 5 Mpc from our Local Group.  They are responsible for the
corresponding overdensity at these distances. However, we did not
include this constrain when we searched for LV-candidates in
simulations for the sake of having better statistics.  Not
surprisingly, the results found from the \LCDM\ and \LWDM\ LVs agree
pretty well among them and with data.

Peculiar motions on the scale of a few Mpc reflect the mass
distribution on these scales and larger ones, hence they should not be
affected by the small scale damping of the primordial perturbations.
We conclude that there is no difference between the two cosmologies in
terms of the internal dynamics of these systems, in good agreement
with other works \citep{coldflow}.

\section{Forming galaxies in haloes}
\label{sec:history}

\begin{figure}
 \begin{center}
  \includegraphics[width=0.5\textwidth]{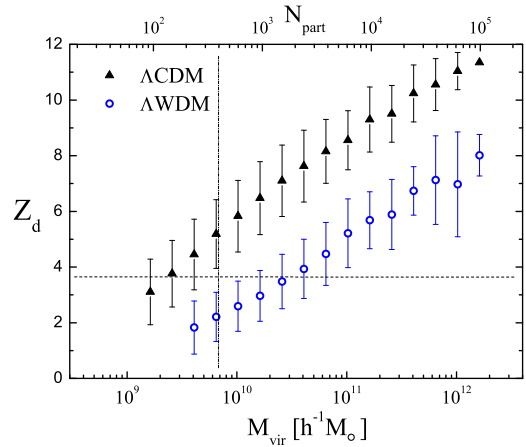}
   \end{center}
   \caption{ Redshift $z_d$ at which the progenitors of present day
      LV haloes  became more massive than $3 \times 10^8 \hMsun$ as a
     function of present day virial mass (upper axis: number of
     particles in the halo). Filled triangles: \LCDM\
     model, open circles: \LWDM\ model.  Dashed lines: see text.}
\label{fig:zm} 
\end{figure}

As we mentioned above, very faint low-mass galaxies have been observed
in the local universe with rotational velocities below $20 \kms$. As
an example, Camelopardalis B with $M_B \sim -10.9$ has a total
dynamical mass of $\approx 10^8 \hMsun$ measured at the last point of
the observed rotation curve \citep{Begum2003}. Due to the strong bias
between dark haloes and galaxies, it is not easy to compare these
observations with collisionless N-body simulations. A proper way of
doing it would be by means of full hydrodynamical simulations which
includes all the relevant physical processes that lead to the
formation of galaxies inside the dark matter potentials. Nevertheless,
we can use results from such kind of simulations to derive estimates
on the formation epochs of the galaxies that would form in the dark
haloes of our simulated local volumes. \cite{Hoeftetal06} have
simulated the formation and evolution of dwarf galaxies in voids using
high resolution  \LCDM\  hydrodynamical simulations  which include radiative
cooling, UV photoionization from a cosmic background, star formation
from the cooled gas and feedback from supernovae.  Based on these
simulations, they derived a characteristic mass scale $M_c(z)$ below
which UV photoheating suppresses further cooling of gas and therefore
star formation inside haloes with $M < M_c(z)$. The numerical results
can be well approximated by a exponentially decreasing function with
redshift in the following way \begin{equation}
   \frac{ M_{\rm c} (z) }{ 10^{10} \: h^{-1} \: M_\odot }
   =
        \frac{0.62 }{(1+z)^{1.23} \exp \left (0.082 z^{2.1} \right)}
        \left\{
                \frac{\Delta_c(0)}{\Delta_c(z)}
        \right\} ^ {1/2}        
\label{mc1}  
\end{equation}
where $\Delta_c(z)$ represents the redshift dependence of the
characteristic virial overdensity of halos that can be obtained from
the numerical integration of the equations of the spherical infall
model or using the analytical approximation obtained by \citet{bryan}.

Thus, by comparing the mass of the progenitors of each of our LV
haloes with $M_c(z) $ we can estimate the typical formation redshift
for the galaxies in our LV's.  Those haloes with mass accretion
histories that are always above $M_c(z)$ have been able to form stars
at all times.  On the contrary, those haloes with mass accretion
histories shallower than $M_c(z)$ would have not been able to condense
baryons inside and they would not host any galaxy. Although these
results were obtained for \LCDM\ halos, in what follows, we will
assume that they can be applied also to our \LWDM\ simulations. Since
the UV photionization background comes from observations and the
modeling of the rest of the other baryonic processes is the same, the
only differences between the \LCDM\ and \LWDM\ halos would come from
the different mass accretion histories.

The "age" of an object in simulations is often determined as the epoch
at which half of its mass is gathered. Thus massive objects form
later.  Contrary to this definition of formation time, there are
others such as the time at which progenitors of present day haloes
reach a given mass threshold. In numerical simulations a natural
threshold is given by the mass resolution which defines a minimum halo
mass as the detection limit. Obviously, more massive objects reached
earlier a given mass threshold and thus, they have formed earlier,
according to this definition of formation time.  Here we use the
detection limit as an estimate for the formation time of the haloes in
the simulation and compare the evolution of \LCDM\ and \LWDM\ haloes
after passing this threshold which we take to be the mass
corresponding to a minimum of 20 particles (i.e $3.2 \times 10^8
\hMsun$).  For the cosmological parameters under consideration the
characteristic mass scale, $M_c(z)$ equals this mass resolution limit
for haloes in our simulations at redshift $z=3.7$ and increases up to
$6.2 \times 10^9 \hMsun$ at redshift $z=0$. Therefore, we are
interested to determine at which redshift $z_d$ the most massive
progenitor of the redshift zero LV haloes reaches the mass threshold
($3.2 \times 10^8 \hMsun$) of our simulations.  To this end, we
identify with an FOF analysis all the progenitors of a given halo at
different redshifts.

In Figure~\ref{fig:zm} we show the mean redshift $z_d$ as a function of
the present day halo's virial mass for all haloes found in the 14 LVs
of the \LCDM\ and \LWDM\ model. We consider that for 
\LCDM\ haloes with more than 100 particles, 
 their merging history can  easily be followed. 
In the case of \LWDM\, we  restrict ourselves to haloes
 above the limiting mass $M_{lim}$ (see \S2). 
 The symbols are centered in the middle of the
selected mass bins and the error bars show the $1\sigma$ scatter. The
difference between the two models is evident: there is a general trend
for \LWDM\ haloes of all masses to reach this threshold later than the
\LCDM\ ones.

The dashed vertical line in Figure~\ref{fig:zm} marks the
characteristic mass for redshift zero, $M_c(0) = 6.2 \times 10^9
\hMsun$.  More massive haloes would have form stars all their lifetime
because their mass accretion histories are always above the $M_c(z)$
since $z_d$ till present. Those haloes with present day masses below
this characteristic mass would be at present practically devoid of
gas due to UV photoheating. Within the \LCDM\ model the
characteristic,  $M_c$,   mass increases faster with time than the mean halo
mass.  Therefore, the mass of those haloes was at some earlier
redshift larger than $M_c(z)$  and thus the haloes were able to form
stars in the past.

The horizontal dashed line in Figure~\ref{fig:zm} marks the redshift
$z^* = 3.7$ at which $M_c(z^*) = 3.2 \times 10^8 \hMsun$ (the
resolution limit). The redshift of the first identified progenitor of
\LCDM\ haloes with $M_{\rm vir} > 2.2 \times 10^9 \hMsun$ is always
above this line. Therefore, all these haloes have been able to cool
gas and form stars at least until redshift 3.7. Those with masses $M
>M_c(0)$ will have formed stars up to present. The smaller haloes,
with $M <M_c(0)$ would have formed stars only until $M_c(z)$ became
larger than the progenitors halo mass (somewhere between $z_d$ and
present). Haloes with $M_{\rm vir} < 2.2 \times 10^9 \hMsun$ lost most
of their gas before $z^* =3.7$. Therefore, they can host only an old
population of stars.

In sharp contrast  with the \LCDM\ case,  the
\LWDM\ haloes with $M_{\rm vir} < M_c(0)$ are always below the
horizontal dotted line. Thus their mass was always smaller than
$M_c(z)$ and they were unable to condense baryons and form galaxies at
all.  This points towards a potential problem of the \LWDM\ model,
namely an insufficient formation of dwarf galaxies.  Since $M_c(z)$
drops down very fast for $z > 4$, in principle they could have been
able to cool gas and form stars at earlier times, resulting in a
population of faint red dwarfs. 

Since haloes form much later in the \LWDM\ scenario they can lose
their gas much easier due to the photoheating of the cosmic  ionizing
background than the \LCDM\ haloes of the same present day
mass. Therefore, the formation of a sufficient number of low mass
galaxies could be a problem in this model. To study this in full
detail, one needs to perform \LWDM\ simulations with much higher mass
resolution than the one presented here.

\begin{figure}
 \begin{center}
  \includegraphics[width= 0.5\textwidth]{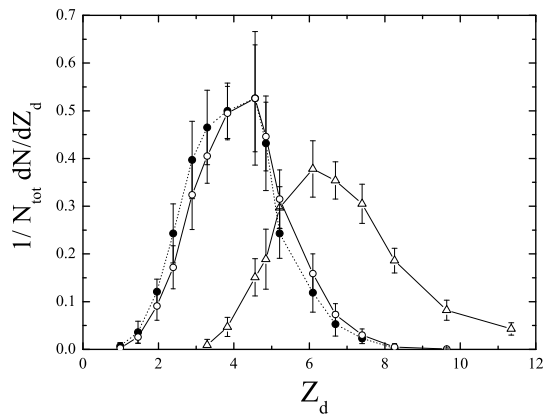}
   \end{center}
   \caption{ The differential number fraction $\frac{1}{N_{tot}}
   \frac{dN}{dz}$ of
   progenitors of haloes which became more massive than $3 \times 10^8
   \hMsun$ at redshift $z_d$. Filled circles: haloes with $V_c < 35
   \kms$ at $z=0$ inside mini-voids, open circles: haloes with $V_c <
   35 \kms$ outside mini-voids, triangles: haloes with $V_c > 35 \kms$. 
   Error bars are $1\sigma$ deviations.}
   \label{fig:nvsz_d} 
\end{figure}

In Figure \ref{fig:nvsz_d} we compare the formation history of \LCDM\
haloes with circular velocities $V_c < 35 \kms $ located inside the
mini-voids (filled circles) with haloes in the same range of circular
velocities, but located outside the voids (open circles). We show the
differential number fraction $\frac{1}{N_{tot}} \frac{dN}{dz}$ of
progenitors of haloes which became more massive than $3.2 \times 10^8
\hMsun$ at redshift $z_d$ which has the same meaning as in in Fig.
\ref{fig:zm}. We used haloes with masses $M > 2 \times 10^9 \hMsun$.
Therefore, the mean $z_d$ of Fig. \ref{fig:nvsz_d} corresponds to the
second filled triangle of Fig. \ref{fig:zm}.  As can be seen in the
figure, regardless of environment, these haloes have very similar
formation histories.  For comparison, we also show the results for the
more massive haloes in the LVs (triangles: $V_c > 35 \kms$) which
obviously passed much earlier the threshold $z_d$. The fact that we do
not see strong differences in the formation epochs of dwarf haloes
located in structures and those sitting in voids may have profound
consequences for the \LCDM\ model.

One of the proposed scenarios to solve the overabundance problem
assumes that UV-photoionization evaporates gas from dwarf haloes in
voids and therefore quenches star formation and leaves voids empty.
If, however, dwarf haloes in voids and the surrounding structures form
roughly simultaneously then photoionization should have suppressed the
formation of all these dwarf galaxies regardless of environment. As
shown above the \LCDM\ model predicts an order of magnitude more
low-mass haloes than observed. If photoionization suppresses the
formation of galaxies in more than 90~\% of the small haloes, it is
difficult to explain why none of the remaining is found in the
voids. Therefore, the observed small number of dwarfs represents a
similar problem for the \LCDM\ model as the missing satellites.

However, the observed dwarfs could also be a potential problem for the
\LWDM\ model.  According to the late formation time of the small mass
haloes they spend almost all their lifetime having masses well below
the characteristic mass $M_c(z)$. Following the arguments of
\cite{Hoeftetal06} it would be difficult to explain how they could
have formed stars at all. However, we should also note that the
characteristic  mass has been derived based on \LCDM\ simulations
\citep{Hoeftetal06,Okamoto2008}.  To get a final answer for halos in
the WDM scenario, proper gas-dynamical simulations including star
formation should be performed. Such simulations should have extremely
high resolution to address also the problem of fake halo formation.

\section{Discussion and Conclusions}\label{sec:concl}

We used an updated version of the \citet{Karachentsev2004} sample of
galaxies to compare the distribution and motions of galaxies in the
Local Volume with that of dark matter haloes in \LCDM\ and \LWDM\
simulations. About 40 mini-voids with sizes in the range from 0.5\,Mpc
to 4\,Mpc have been detected in the observational sample. \cite{TK08}
have shown that the spectrum of void sizes is stable with respect to
variations of the sample and to uncertainties of distances to
individual galaxies. 

From our \LCDM\ and \LWDM\ simulations we have selected the Local
Volume candidates following the criteria used by \cite{TK08}.  In many
respects these LV-candidates look very similar to the observed Local
Volume: they have comparable number of bright (massive) objects and a
similar density contrasts on 8Mpc scale. Moreover, most of the
LV-candidates show the same typical intersecting filaments sometimes
nearly aligned in planes which are quite similar to what one observes
in the Local Volume - intersecting filaments form a structure which is
part of the Local Supercluster and goes through the Local Group
(sometimes it is called "Local Sheet"). The 14 Local Volume candidates
in \LCDM\ and \LWDM\ simulations show only little scatter in the
volume fraction occupied by mini-voids. Thus it  is well suited to
discriminate between different models. To test further our
LV-candidates we have calculated the rms radial velocity deviations
from the Hubble flow and found no difference in the dynamics of the
haloes. Both in the \LCDM\ and \LWDM\ models $\sigma_H$ is in
agreement with the observations.
 
In Sect. \ref{sec:vmax} we argue in favor of an association of the
haloes maximum circular velocity to the observed rotational
velocity. Our argument is based on a comparison of the rotational
velocities observed in a certain distance from the center and the
circular velocities of dark matter haloes in the same distance. The
velocity profiles of the dark matter haloes in are essentially flat at
the distance where the rotational velocities of dwarfs are
measured. Thus the errors are small if one associates the measured
rotational velocity to the peak circular velocity. The exact relation
between both depends on the gasdynamical processes in the galaxy which
we cannot consider in the large volume under discussion. One might
argue that the peak velocity increases due to adiabatic contraction so
that the WDM model would be even more favored. However, these dwarf
haloes are expected to be dark matter dominated and adiabatic
contraction is expected to be less important. One might argue that by
some processes the rotational velocity in the dwarf haloes becomes
smaller but we cannot see a good reason to associate $V_{rot}$ with
significantly more massive haloes which would be necessary to fit the
observational data.

Within the \LCDM\ model the cumulative volume fraction of the LV
occupied by mini-voids matches observations only if one assumes that
objects with $V_{c} > 35 \kms $ define the local mini-voids and these
objects host galaxies brighter than $M_B = -12$ \citep{TK08}. The
\LCDM\ model predicts almost 500 haloes with $20 < V_c < 35 \kms$
within the mini-voids in a LV. At present 10 quite isolated dwarf
galaxies have been observed in the Local Volume with magnitudes
$-11.8 > M_B > -13.3$ and rotational velocities $V_{rot} < 35 \kms$
\citep{TK08}. This number is more than an order of magnitude smaller
than the predicted number of low-mass haloes and thus it points to a
similar discrepancy as the well known predicted overabundance of
satellites in Milky Way sized \LCDM\ haloes. The reason can be
understood from the $\Lambda$CDM mass function at the low-mass end
which is much steeper than the observed luminosity function at the low
luminosity end. This causes both the overabundance of satellites and
the fact, that the observed mini-void distribution can be explained
only if haloes with circular velocities larger than $35\kms$ host
galaxies. In case of satellites one could explain the mismatch between
observations and predictions by processes like ram-pressure and tidal
stripping which have not been taken correctly into account in
simulations. On the contrary, the dwarfs in the Local Volume are all
situated in the same environment of relatively low density. Therefore,
the existence of only a small number of such low luminosity and low
mass dwarf galaxies in the LV is even more difficult to explain than
the missing satellites.

In case of the \LWDM\ model the observed spectrum of mini-voids can be
explained naturally if dark matter haloes with circular velocities
larger than $\sim 15-20 \kms$ host galaxies. The observed small number
of  slowly rotating isolated galaxies is not in contradiction with
this assumption since a small number of low-mass haloes are also  formed
 in the \LWDM\ model. Thus the \LWDM\ model solves the problem of
overabundance and the spectrum of mini-voids in a distribution of
low-mass haloes is in agreement with observations. However, we have
also shown that due to the late formation of haloes in the \LWDM\
model a substantial part, if not all, of the low-mass haloes could have
failed to form stars. Thus they would have not been 
 able to host any visible galaxy, as it is necessary for the model to
 agree with the sizes  of mini-voids in the Local Universe.
Nevertheless, to study their distribution and properties in detail one
needs simulations with much higher resolution so that one can
distinguish between the real low-mass haloes and the numerical
artifacts. Moreover, full radiative hydrodynamical simulations with
star formation are also needed in order to properly address the galaxy
formation process in these halos. Overall we confirm earlier claims
that concerning the formation of small scale structure the \LWDM\
model can be considered as a promising alternative to the \LCDM\
model. However, more detailed investigations are necessary to study to
which extend star formation in low-mass haloes is suppressed in the
\LWDM\ model.

\section*{Acknowledgments}
We thank I.D.\,Karachentsev for providing  us  with an updated list of his
Catalog of Neighboring Galaxies.  The simulations used in this work 
were performed at the Leibniz Rechenzentrum Munich (LRZ) and at  the Barcelona
Supercomputing Centre (BSC), partly granted by the DEISA Extreme Computing
Project (DECI) SIMU-LU.  This work was funded by the Deutsche
Forschungsgemeinschaft (DFG grant: GO 563/17-1) and supported by the
ASTROSIM network of the European Science Foundation (ESF). SG
acknowledges a Schonbrunn Fellowship at the Hebrew University
Jerusalem. YH has been partially supported by  
the ISF (13/08).  GY acknowledges support of the Spanish Ministry of
Education through research grants FPA2006-01105 and AYA2006-15492-C03.
We thank A.\,Klypin for useful discussions and suggestions.

\bibliographystyle{mn2e}

\label{lastpage}

\end{document}